\documentclass{ws-procs9x6}

\begin{document}

\title{PARITY-VIOLATING NUCLEON-NUCLEON\\ INTERACTIONS:
 WHAT CAN WE LEARN\\ FROM NUCLEAR ANAPOLE MOMENTS?}

\author{B. DESPLANQUES$^*$ }

\address{LPSC, Universit\'e Joseph Fourier Grenoble 1, CNRS/IN2P3, INPG,\\
  F-38026 Grenoble Cedex, France \\
$^*$E-mail: desplanq@lpsc.in2p3.fr}

\begin{abstract}
Knowledge about parity-violating effects both from theory and experiment is
reviewed. Further information that could be obtained from measurements of
nuclear anapole moments is discussed.
\end{abstract}

\bodymatter
\section{Introduction} 
In the present contribution, we are concerned with hadronic weak interactions
and, more specifically, with the strangeness-conserving component. This one,
contrary to the strangeness-violating one, is masked by strong (and
electromagnetic) interactions. It can be disentangled through its parity-violating
(pv) component, which leads to effects of order $10^{-7}$ for low-energy $NN$
interactions of interest here. Its experimental study is therefore expected
to be difficult. It is however necessary to complement the knowledge from
strangeness-violating processes. Until now, only the sector 
of nuclear interactions has been the object of experimental studies. 

Similarly to the $NN$ strong interaction, the pv interaction is
supposed to occur at the two-nucleon level in first place.
Various approaches have been considered with different emphasis depending 
on time: more ambitious ones, based on meson exchange 
(see Refs. \cite{DDH_ap80,km_npa89} for instance), 
alternating with more phenomenological ones, 
devoted to the consistency of effects observed at low
energy \cite{danilov_sjnp72,missimer_prc76,desp_npa78,zhu_npa05}. 
In any case, the description of pv $NN$ interactions involves a minimal
set of 5 pieces of information corresponding to the pv elementary $S-P$
transition amplitudes.

Looking at nuclear pv effects observed up to now, it is found that, 
for a large part, they individually agree with expectations 
within a factor~2~\cite{adel_arnp85,desp_phrep98}. 
This is not however sufficient to get a consistent description 
and, thus, a reliable determination of pv  $NN$ forces. 
Further studies are required to provide the missing information. 
With this respect, we intent to review here some benchmark results, 
make comments partly in relation with recent works and  
show in what measurements of nuclear anapole moments could be useful. 
We largely refer to Ref. \cite{desp_phrep98} for omitted details.
 
The present paper is organized as follows. In Sec. 2, we review 
the description of the pv $NN$ forces in terms of meson exchanges. 
This includes a discussion of the numerous uncertainties pertinent 
to this approach. Section 3 is devoted to the phenomenological approaches 
considered in the past and recently. A possible hint for the failure 
of the single-meson exchange picture is described in Sec. 4. 
Presently known information is reviewed in Sec. 5.
The conclusion describes in what the  knowledge of nuclear anapole moments
could be useful for getting a better determination of pv  $NN$ forces.
\section{Meson-exchange description of pv $NN$ forces} 
\begin{figure}[htb]
\begin{center}
\mbox{ \psfig{ file=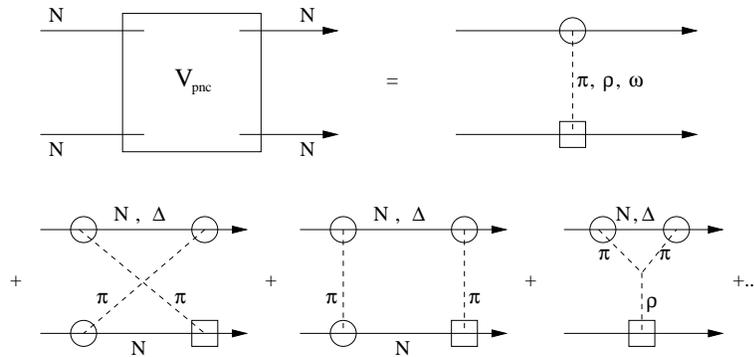, width=10cm}}
\end{center}
\caption{Graphical representation of single-meson and some two-meson
exchanges\label{fig:pot} }
\end{figure}  
Diagrams representing contributions to the pv $NN$ forces considered 
up to now are shown in Fig. \ref{fig:pot}. They involve the exchange 
of a single meson ($\pi,\; \rho,\;\omega$) where the pv vertex (squared box) 
is described by a constant to be determined while the circle represents 
the strong-interaction vertex, in principle known. As the weak interaction 
does not conserve isospin, many couplings are possible in some cases. 
They are usually denoted as:\\
$h^1_{\pi}:\;\;\;\; \Delta T=1$ (long range, {\it a priori} favored), \\
$h^{1'}_{\rho}:\;\;\; \Delta T=1$ (short range),\\
$h^{0,1,2}_{\rho}: \Delta T=0,1,2$ (short range),\\
$h^{0,1}_{\omega}:\;\; \Delta T=0,1$ (short range).\\
The  $\pi^0$ contribution is absent (Barton's theorem \cite{barton_nc61}) 
as well as that one due to other spin-0 mesons such as $\eta^0$, $\sigma^0$ 
(or two pions in a S state \cite{chem_npa74}). 

Two-pion exchange contributions shown in Fig. \ref{fig:pot} were considered 
in the~70's,  within a covariant approach 
\cite{pign_plb71,desp_plb72,pirner_plb73,chem_npa74}. Some were recently considered 
at the dominant order within an effective field-theory framework. The dots 
in Fig. \ref{fig:pot} stand for contributions due to heavier-meson, 
multi-meson or excited-baryon  exchanges. These contributions, 
which are expected to have a quite short range, were discarded in the 70's 
on the basis  of a large repulsion in the $NN$ strong interaction models
at these distances. This feature appears to be now a consequence of the local
character of the models then used \cite{amghar2_npa03}. The effect is much less pronounced 
with non-local models such as CD-Bonn \cite{mach_prc01} 
or some Nijmegen ones \cite{stoks_prc94}. 
As a consequence, the support to neglect the above short-range 
pv contributions is now much weaker. On the other hand, 
these extra contributions could involve new parameters, 
making more complicated the description of pv  $NN$ forces. 
One can only hope they are not too large. 

From examining $NN$ strong interaction models, there are hints that 
the approximation of a single-meson exchange should be considered with caution. 
It has thus been shown that the combined effect of $\pi$ and $\rho$ exchange 
could provide some sizable repulsion \cite{durso}. In its absence, the $\omega NN $ coupling
has to be increased from an expected value of $g^2_{\omega NN}/4\pi\simeq 9$ 
to a value of about 20. On the other hand, a model like Av18 \cite{wir_prc95}, 
which, apart from one-pion exchange, is entirely phenomenological, 
fits experiments with a $\chi^2$ of 1 {\it per data}. 
This shows that the relevance of a single-meson exchange 
is far to be established. Thus, apart from $h^1_{\pi}$, pv couplings 
entering a single-meson exchange model based on
$\pi,\; \rho,\;\omega$ could be quite effective. 

Estimates for pv meson-nucleon couplings have been made by many authors.
Most of them can be shown to be part of the DDH ones \cite{DDH_ap80}. 
We therefore rely on this work for a discussion. 
Some detail is given in Table~1 of Ref. \cite{desp_phrep98}, 
the most important couplings being $h^1_{\pi},\; h^0_{\rho},\; h^0_{\omega}$.  
They are obtained from an effective $qq$ interaction, which depends on the factor 
$K\simeq 1+\frac{25}{48\,\pi^2}\; g^2(\mu)\; {\rm log}\Big(
\frac{M^2_W}{\mu^2}\Big)$, which accounts for QCD strong-interaction effects. 
Part of the contributions is based on the
$SU(6)_W$ symmetry and knowledge from non-leptonic hyperon decays.  
Another part is based on a factorization approximation. The original range 
assumed $ 1\leq K \leq 6$, possible $SU(6)_W$ symmetry breaking and some weight 
for the neutral-current contribution. The best-guess values were obtained  by
weighting the various ingredients, in particular with respect 
to $SU(6)_W$ symmetry breaking and $q\bar{q}$-pair role. The update, 
which takes into account the fact that some ingredients are better known now, 
mainly affects the coupling $h^1_{\pi}$ (see Table~4 of Ref.~\cite{desp_phrep98} ). 
Results from a quite different approach based 
on a soliton model (KM) \cite{km_npa89} can also be found 
in Table~1 of Ref. \cite{desp_phrep98}. Instead, results for $h^1_{\pi}$ 
obtained from QCD sum rules \cite{khat_sjnp85,henl_plb96,hwang_zphys97} 
are not shown.

Examination of various results suggests many remarks. The comparison 
of DDH and KM results evidences interesting similarities despite differences 
in the approaches.  In both cases, the contribution to $h^1_{\pi}$  
due to non-strange quarks is suppressed, 
most of the estimate involving strange quarks. The estimates for 
$ h^0_{\rho}$ and $h^0_{\omega}$ differ but rough agreement could be obtained 
by weighting differently the various contributions in DDH (notice that KM
provides couplings at $q^2=0$ while DDH give them at the meson pole). On the
basis of DDH estimates, the contribution of $h^0_{\omega}$ has often been
neglected but KM results show it should not be. On the other hand, the upper
limit for this coupling in DDH has not been reproduced. Its range 
should be mostly negative. As for the  $h^1_{\pi}$ estimates from QCD
sum rules \cite{khat_sjnp85,henl_plb96,hwang_zphys97}, 
we notice that, as far as we can see, they do not correspond 
to any of the expected results shown in Table~4 of Ref.~\cite{desp_phrep98}). 
They are obtained 
from non-strange quarks only and involve a coherent sum 
of two contributions while cancellations are observed in DDH and KM. 
It is not clear at the present time whether they represent a new
contribution or whether their relation to other results has not
been found yet.  

It has been mentioned that predictions for couplings should be corrected for
various effects \cite{zhu2} (rescattering, vertex corrections). 
These ones could
affect the factorization part of DDH estimates but should not change 
the other parts which, being based on experimental non-leptonic 
hyperon decays, already account for them. 
The problem in this case is whether $SU(6)_W$ symmetry breaking effects 
are correctly accounted for. 
\section{Phenomenological approaches to pv $NN$ forces} 
For a part, phenomenological approaches have been motivated by the failure 
of single-meson-exchange approaches to provide a consistent description of 
various data. Less ambitious than the potential ones, they were originally 
limited to the five $S-P$ transition amplitudes 
\cite{danilov_sjnp72,missimer_prc76}. 
The important point is that their energy dependence in the few MeV range 
 is completely determined by well-known strong-interaction 
properties. Thus, five parameters are required for their description 
\cite{danilov_sjnp72,missimer_prc76}. The approach was extended 
to higher energies, motivated by the fact that most effects known 
at that time were involving complex nuclei \cite{desp_npa78}. 
In this aim, a 6th parameter, 
chosen as the $h^1_{\pi}$ coupling, was added. Due to its long range, 
the $\pi$-exchange force induced by this coupling indeed provides 
an extra sizable e\-ner\-gy dependence for the $^3S_1 -\,\! ^3P_1$ 
transition amplitude beyond a few MeV.  Moreover, it contributes significant
$^3P_{1,2}-\,\! ^3D_{1,2}$ transitions in complex nuclei, 
with a sign opposite to the $^3S_1-\,\! ^3P_1$ one. 
The  configuration-space expression of the interaction 
can be found in Refs. \cite{desp_npa78,desp_phrep98}. 

The approach was considered again recently but within the framework 
of an effective-field theory \cite{zhu_npa05}. This one involves a chiral expansion in terms 
of a quantity $\Lambda_{\chi}^{-1}$. The number of parameters to describe 
the pv $NN$ interaction is the same as above. 
However, the part of two-pion exchange induced by the $h^1_{\pi}$ coupling 
at the first possible order, $\Lambda_{\chi}^{-3}$, is separated out 
with the motivation it is a medium-range interaction.
At the same order, a specific contribution to the electromagnetic
interaction appears, hence a 7th parameter in the approach.
The momentum-space expression of the two-pion exchange contribution 
can be found in Refs. \cite{zhu_npa05,desp_prc08}. 
It only contains two terms that correspond 
to the local ones of a more ge\-ne\-ral expression considered in the 70's 
\cite{pign_plb71,desp_plb72,pirner_plb73,chem_npa74}. 
Apart from some mistakes in the original work, 
the comparison with earlier works shows that 
\cite{desp_prc08}: i) the leading-order approximation
overestimates the results at finite distances, ii) the non-local terms that
appear at the next order are not negligible, iii) in comparison to a $\rho$ 
exchange, the two-pion exchange has a longer range as expected 
from the two-pion tail but also a shorter range. 
In this case, the result is due to the fact that
the two pions are exchanged in a P state. The exchange in a S state, 
which dominates at intermediate distances in the strong-interaction case, 
does not contribute here  \cite{chem_npa74}.

For the purpose of analyzing experimental data, we introduce dimensionless
quantities which are more closely related to the isospin properties of the
system under consideration \cite{desp_npa78}. They are $X_{pp}$ and $X_{nn}$, which involve 
the $pp$ and $nn$ forces, and $X^+_{pn}(+5.5h^1_{\pi}) $, 
$X^-_{pn}(-5.5h^1_{\pi}) $, $X^0_{pn}$, which involve
the $pn$ force. The strengths $X^+_{pn}$ and $X^-_{pn}$ are appropriate for the
description of the $pn$ force in odd-proton- and odd-neuton-nucleus forces 
while $X^0_{pn}$ does not contribute to them. The $\pi$-exchange contribution,
is put between parentheses for the case it would be considered explicitly. We
also introduce the combinations:
$X^p_N=X_{pp}+X^+_{pn}(+5.5h^1_{\pi}) $, 
$X^n_N=X_{pp}+X^-_{pn}(-5.5h^1_{\pi}) $,
which determine the strengths of the proton- and neutron-nucleus forces (for a
nucleus with spin and isospin equal to 0).

The $X $ parameters are closely related to the Danilov ones \cite{desp_npa78}. 
We stress that they account for a lot of short-range unknown physics 
($NN$ short-range correlations, heavy- and multi-meson exchange, 
relativity, $\cdots$). If necessary, they can be calculated in terms 
of the $h$ couplings entering  a meson-exchange description 
of the pv $NN$ interaction, allowing one to check the underlying model
\cite{desp_phrep98,liu_xxx06}.
\section{Possible failure of the single-meson exchange model} 
It has been proposed that  the study of pv effects in $pp$ scattering 
at different energies could allow one to disentangle contributions 
due to $\rho$ and $\omega$ exchanges. There are measurements 
at 13.6 MeV \cite{ever_plb91} and 45 MeV \cite{kist_prl87},
which mainly involve a $^1S_0-\,\!^3P_0$ transition, 
and at 221 MeV \cite{berdoz_prl01}, which is sensitive to
the $^3P_2- \,\!^1D_2$ transition. The analysis has been performed 
by Carlson et al., with the results \cite{carlson_prc02}:
$10^7\;h^{pp}_{\rho}=-22.3,\;\;10^7\;h^{pp}_{\omega}=+5.2$.\\
The $\rho$ coupling agrees with DDH expectations while the $\omega$ one 
is at the extreme limit of the proposed range but disagrees if one notices 
that the range is now restricted to negative values. 
In short, it is found that the contribution to the 
$^3P_2- \,\!^1D_2$ transition is too small, roughly by a factor 2. 
It has been thought that a longer-range force could enhance 
the contribution to the $^3P_2- \,\!^1D_2$ transition relatively 
to the $^1S_0-\,\!^3P_0$ one \cite{liu_prc05}. 
While this is verified for the undistorted Born amplitude, it is not for the
distorted one. Due to the effect of a strong short-range repulsion 
for the $^1S_0-^3P_0$  transition amplitude, 
the effect is the other way round, making the problem more severe.
There are different issues: i) the measurement at 221 MeV is too large (by a
factor of about 2), ii)  estimates of pv coupling constants miss important
contributions, raising some doubt about present ones, 
iii) besides the single-meson exchange contribution, there are 
important extra contributions (multi-meson exchange, $\cdots$). For this last
issue, there would be no other choice than to adopt an effective approach, in
which case, the $^3P_2- \,\!^1D_2$ and $^1S_0-\,\!^3P_0$ transition amplitudes are
described by independent parameters.

\section{Presently known information} 
The first information we want to consider concerns the strength 
of the pv $pp$ force, $X_{pp}$. It can be obtained from measurements 
of the pv asymmetries $A_L$ in $pp$ scattering at  13.6 and 45 MeV 
($A_L=-0.10\; {\rm and}\; -0.17X_{pp}$ respectively \cite{sim_plb72}). 
The fitted value \cite{desp_phrep98}:
\begin{equation}
X_{pp}\approx 0.9\times 10^{-6}, \label{xpp}
\end{equation}
provides a very good description of the measurements 
\cite{kist_prl87,ever_plb91}, leaving little space 
for a $^3P_2- \,\!^1D_2$ transition, which is expected to be small in any case. It is
stressed that the coefficients 0.10 and 0.17 are determined by well-known
properties of the strong $NN$ interaction, while the unknown physics is
incorporated in the quantity $X_{pp}$. The above value provides an unambiguous 
benchmark for the strength of pv $NN$ forces.

The second information is obtained from the strength of the proton-nucleus
force, $X_N^p$, mostly determined by pv effects observed in odd-proton systems 
such as $p-\alpha$ scattering or radiative transitions in complex nuclei 
($^{19}F,\;^{41}K,\;^{175}Lu,\;^{181}Ta $) \cite{desp_phrep98}. While the dependence 
of most effects on  $X_N^p$ results from the underlying nuclear model 
used in the estimate \cite{desp_npa79}, this is not so in $^{19}F$ 
where the estimate was based on a shell-model calculation \cite{adel_prc83}. 
Examination of the detailed calculations in the last case nevertheless 
shows that the result has the structure of a single-proton transition in an
average field determined by the strength $X_N^p$, evidencing the role 
of two effects that were accounted for in the heavier nuclei 
(pairing and deformation). Another calculation  \cite{brown_prl95} 
however suggests that correlations could affect differently 
isovector and isocalar contributions to $X_N^p$ (see also below). 
A good description 
of the observed effects assumes the fitted value \cite{desp_phrep98}: 
\vspace*{-0.5mm}
\begin{equation}
X_N^p \approx 3.4 \times 10^{-6}. \label{xp}
\vspace*{-0.5mm}
\end{equation}
Examining the whole fit (within experimental errors), we are inclined 
to think it is too good however. At this point, we simply notice that the
contribution of the $pp$ force to the proton-nucleus force is relatively small,
pointing to a large contribution of the $pn$ force which could be due either to
the isoscalar part of the pv force or to the isovector one. 

The third available information concerns this isovector part of the force. 
It is obtained from the analysis of pv effects in the transition 
$^{18}F(0^- \rightarrow1^+)$. An extensive analysis of the effect 
in this process has been done \cite{adel_prc83}. 
It essentially involves the difference of the pv proton- and
neutron-nucleus forces and results in the following upper limit:
\vspace*{-0.5mm}
\begin{equation}
|X_N^p-X_N^n| \leq 1.4 \times 10^{-6}, 
\vspace*{-0.5mm}
\end{equation}
from which we can derive a range relative to the neutron-nucleus force: 
\vspace*{-0.5mm}
\begin{equation}
 2.0\;(2.6)\times 10^{-6} \leq X_N^n  \leq 4.2\;(4.8)\times 10^{-6} .
\vspace*{-0.5mm}
\end{equation}
Assuming that the isovector contribution is dominated by the single-pion
exchange force, one would get: 
$|h_{\pi}^1| \leq 1.3 \times 10^{-7}$. To some extent, the absence 
of effect in $^{21}Ne$ supports the above limit, which could even be smaller
if the isoscalar contribution tends to be suppressed \cite{desp_npa93}. 
To a lesser extent, results for $^{75}Tc$  \cite{hass_plb96} 
lead to a similar conclusion \cite{desp_phrep98}.

Concerning the other pieces of information, $X_{nn}$ and $X^0_{np}$, 
one could rely on little details in the theoretical estimates 
to determine them from the analysis of pv effects in various processes. 
The comparison of these estimates, especially in complex nuclei, 
however shows that these details are somewhat uncertain, 
preventing one to get reliable information. The strength $X_{nn}$ would be
best determined from measurements involving neutrons in light systems. 
The  strength  $X^0_{np}$ does not play much role in complex nuclei 
\cite{desp_phrep98}. The  most
favorable process for its determination is the measurement of the photon
circular polarization in the radiative capture $n+p \rightarrow d+\gamma$. 
Determining the 6th parameter, $h_{\pi}^1$, introduced to get 
a better description of the pv $NN$ interaction at low energy, 
supposes to dis\-en\-tangle its long-range contribution from a short-range one. 
As the first contribution is expected to dominate the other one however, 
$h_{\pi}^1$ could be best determined from the measurement of the pv asymmetry 
in $\vec{n}+p \rightarrow d+\gamma$ where the effect is maximized 
(see Refs. \cite{desp_phrep98,hyun_plb07,desp_prc08} and references therein).
Determining the short-range part could be quite difficult in practice.

Though the information is incomplete, one can nevertheless 
have an interesting discussion relative to the strength 
of the $pp$ force, $X_{nn} \simeq 0.9 \times 10^{-6}$ and its contribution 
to the proton-nucleus one, $X_N^p \approx 3.4\times 10^{-6} $. 
The re\-la\-tive sign is encouraging but the relative size supposes that a large
contribution to $X_N^p$ comes from a $pn$ force. This is questionable 
in absence of a large pion-exchange contribution, as constrained 
from $^{18}F(0^- \rightarrow1^+)$. In usual potentials models,
the isoscalar $pn$ contribution to $X_N^p$ is at best of the order 
of the $pp$ one. Thus, its  strength could be larger than expected by a factor
from 2 to 3.
This failure could indicate that the usual potential models miss some
contribution, supporting for a part findings from $pp$ scattering.
Interestingly, the analysis of pv effects in this process (see Sec.~4), 
with a positive 
$\omega NN$ coupling, tends to enhance the strength of the $pn$ force with
respect to the $pp$ one \cite{carlson_prc02,liu_prc05}. 
Another explanation supposes medium effects that could
enhance the strength of the proton-nucleus pv force in heavy nuclei. 
Some mechanisms, in relation with an attraction in the isoscalar 
$0^{-}$ channel \cite{flam_prc94,brown_prl95,desp_phrep98}  (RPA correlations) 
or relativity \cite{horo_prc94}, have been discussed in the literature  
but the size of the effect depends on poorly known ingredients.
\section{What from nuclear anapole moments? Conclusion} 
In first approximation, nuclear anapole moments involve the strengths of the
proton- and neutron-nucleus forces, $X_N^p$ and $X_N^n$ 
\cite{flam_jetp80}. Until now, there is no
direct determination of the last one. Measurements of anapole moments in
odd-neutron nuclei could therefore provide a valuable information on the
strength $X_N^n$.  This information could also be obtained from pv effects
in \mbox{$n-\alpha$} scattering (see Ref.~\cite{desp_wsc00} 
and references therein), unless there are sizable medium effects. In such a
case, the two measurements will complement each other and their comparison
could allow one to determine the size of these corrections for which there is
some hint in odd-proton nuclei.

The measurement of anapole moments in odd-proton nuclei has already been
performed. In  the most accurate case however  ($^{133}Cs$ \cite{wood_sci97}), 
the strength of $X_N^p$  required to account for the measurement is roughly 
twice as much as that one given in Eq. (\ref{xp}), 
obtained from other odd-proton systems \cite{dmit_npa94,liu_prc02}. 
A factor 2 is typical of theoretical 
nuclear uncertainties in estimates of pv effects in nuclei but, 
looking at different calculations, it sounds that the uncertainty 
is smaller in the case of $^{133}Cs$,  hence some serious concern. 
Noticing that the strength of the proton-nucleus force could be larger than
expected on the basis of its contribution due to the $pp$ force 
(see previous section), 
the result in $^{133}Cs$ could simply be explained by an enhancement 
effect that is already at work in other complex nuclei. 
If so, one could wonder why effects in heavy nuclei have not required 
such an effect, especially in $^{41}K$ and $^{175}Lu$, 
where previous calculations were relatively stable. 
An explanation could be as follows. These calculations 
assumed that initial and final states are described by very simple 
configurations with some correlations (pairing, deformation) preserving 
the single-particle character of the pv transition. 
More detailed calculations, made later on in lighter nuclei 
on a similar basis, have shown that the weight of such a contribution
is often  decreased by the consideration 
of further correlations \cite{adel_prc83}. 
In this case, some enhancement of the strength of the proton-nucleus force
could also be required in $^{41}K$ and $^{175}Lu$. Having shown that the
anapole moment in $^{133}Cs$ does not necessarily contradict other pv effects
in nuclei, provided that some medium effect is invoked, 
we believe that the measurement of anapole moments in other odd-proton 
complex nuclei could be helpful in clarifying the present understanding 
of pv effects in complex nuclei.

By looking at anapole moments of nuclei with different numbers of protons 
and neutrons, one could imagine to also determine the separate contributions
to $X_N^p$  ($X_N^n$) due to $X_{pp}$ ($X_{nn}$) and $X^+_{pn}$ ($X^-_{pn}$).
Involving smaller contributions, this program would however suppose 
that both measurements and theoretical estimates are very accurate. 
This could be an interesting program for a much further future, 
once strengths, $X_N^p$ and  $X_N^n$, are unambiguously determined.


\begin{thebibliography}{99}

\bibitem{DDH_ap80}
 B. Desplanques, J. F. Donoghue and B. R. Holstein,
{\it Ann. Phys. (N.Y.)} \textbf{124}, 449 (1980).

\bibitem{km_npa89}
N. Kaiser and U.-G. Meissner, {\it Nucl. Phys. A} \textbf{499}, 699 (1989).

\bibitem{danilov_sjnp72}
G. S. Danilov, {\it Sov. J. Nucl. Phys.} \textbf{14}, 443 (1972).

\bibitem{missimer_prc76}
J. Missimer, {\it Phys. Rev. C} \textbf{14}, 347 (1976).

\bibitem{desp_npa78}
B. Desplanques and J. Missimer, {\it Nucl. Phys. A} \textbf{300}, 286 (1978). 

\bibitem{zhu_npa05}
S.-L. Zhu {\it et al.}, {\it Nucl. Phys. A} \textbf{748}, 435 (2005).

\bibitem{adel_arnp85}
E.  Adelberger and W.  Haxton, {\it Ann. Rev. Nucl. Part. Sci.} \textbf{35}, 501
(1985).

\bibitem{desp_phrep98}
B. Desplanques, {\it Phys. Rept.} \textbf{297}, 1 (1998).

\bibitem{barton_nc61}
G. Barton, {\it Nuovo Cimento} \textbf{19}, 512 (1961).

\bibitem{chem_npa74} 
M. Chemtob and B. Desplanques, {\it Nucl. Phys. B} \textbf{78}, 139 (1974).

\bibitem{pign_plb71}
D. Pignon,  {\it Phys. Lett.} \textbf{35}B, 163 (1971).

\bibitem{desp_plb72} 
B. Desplanques, {\it Phys. Lett.} \textbf{41}B, 461 (1972).

\bibitem{pirner_plb73} 
H. Pirner and D. O. Riska, {\it Phys. Lett.} \textbf{44}B, 151 (1973).

\bibitem{amghar2_npa03} 
A. Amghar and B. Desplanques, {\it Nucl. Phys. A} \textbf{714}, 502 (2003).

\bibitem{mach_prc01}   
 R. Machleidt, {\it  Phys. Rev. C} \textbf{63}, 024001 (2001).

\bibitem{stoks_prc94} 
V. G. J. Stoks {\it et al.}, {\it Phys. Rev. C} \textbf{49}, 2950 (1994).

\bibitem{durso}
J. W. Durso \textit{et al.}, {\it Nucl. Phys. A} \textbf{278}, 445 (1977). 

\bibitem{wir_prc95} 
R. B. Wiringa, V. G. J. Stoks and R. Schiavilla, 
   {\it Phys. Rev. C} \textbf{51}, 38 (1995).

\bibitem{khat_sjnp85} 
V. M. Khatsimovskii, {\it Sov. J. Nucl. Phys.} \textbf{42}, 781 (1985).

\bibitem{henl_plb96} 
E. M. Henley {\it et al.}, {\it  Phys. Lett. B}  \textbf{367}, 21   (1996).

\bibitem{hwang_zphys97} 
W. H. P. Hwang, {\it Z. f\"ur Physik C} \textbf{75}, 701 (1997).

\bibitem{zhu2}
S.-L. Zhu \textit{et al.}, {\it Phys. Rev. D} \textbf{63}, 033006 (2001).

\bibitem{desp_prc08}
B. Desplanques {\it et al.}, nucl-th/0803.2075.

\bibitem{liu_xxx06} 
C.-P. Liu, {\it Phys. Rev. C} \textbf{75}, 065501 (2007).

\bibitem{ever_plb91}
P. D. Eversheim \textit{et al.}, {\it Phys. Lett. B} \textbf{256}, 11 (1991). 

\bibitem{kist_prl87}
S. Kistryn \textit{et al.}, {\it Phys. Rev. Lett.} \textbf{58}, 1616 (1987). 

\bibitem{berdoz_prl01}
A. R. Berdoz \textit{et al.}, {\it Phys. Rev. Lett.} \textbf{87}, 272301
(2001). 

\bibitem{carlson_prc02}
J. Carlson {\it et al.}, 
{\it Phys. Rev. C} \textbf{65}, 035502 (2002). 

\bibitem{liu_prc05}
C.-P. Liu, C. H. Hyun and B. Desplanques, {\it Phys. Rev. C} \textbf{73}, 065501 (2006).

\bibitem{sim_plb72}
M. Simonius, {\it  Phys. Lett.} \textbf{41}B, 415 (1972), 
{\it Nucl. Phys. A} \textbf{220}, 269 (1974). 

\bibitem{desp_npa79}
B. Desplanques, {\it Nucl. Phys. A} \textbf{316}, 244 (1979).

\bibitem{adel_prc83}
E.  Adelberger \textit{et al.},  {\it Phys. Rev. C} \textbf{27}, 2833 (1983).

\bibitem{brown_prl95}
M. Horoi and B. A. Brown, {\it Phys. Rev. Lett.} \textbf{74}, 231 (1995). 

\bibitem{desp_npa93}
B. Desplanques and O. Dumitrescu, {\it Nucl. Phys. A} \textbf{565}, 818 (1993). 

\bibitem{hass_plb96}
M. Hass \textit{et al.}, {\it   Phys. Lett. B} \textbf{371}, 25 (1996).

\bibitem{hyun_plb07} 
C. H. Hyun, S. Ando and B. Desplanques, {\it  Phys. Lett. B} \textbf{651}, 257 (2007).

\bibitem{flam_prc94}
V. V. Flambaum and O. K. Vorov, {\it Phys. Rev. C} \textbf{49}, 1827 (1994).

\bibitem{horo_prc94}
C. J. Horowitz and O. Yilmaz, {\it Phys. Rev. C} \textbf{49}, 3042 (1994).

\bibitem{flam_jetp80}
V. V. Flambaum and I. B. Khriplovich, {\it Sov. Phys. JETP} \textbf{52}, 835
(1980).

\bibitem{desp_wsc00}
B. Desplanques, in {\it   Fundamental Physics with Pulsed Neutrons Beams
(FPPNB-2000)}, eds. C. R. Gould \textit{et al.} (World Scientific, 2000) 
pp. 87--96.

\bibitem{wood_sci97}
C. S. Wood \textit{et al.}, {\it Science} \textbf{275}, 1759 (1997).

\bibitem{dmit_npa94} 
V. F. Dmitriev \textit{et al.},  {\it Nucl. Phys. A} \textbf{577}, 691 (1994). 

\bibitem{liu_prc02} 
W. Haxton \textit{et al.}, {\it Phys. Rev. C} \textbf{65}, 045502 (2002).


\end{thebibliography}

\end{document}